# Moral Foundations of Political Discourse: Comparative Analysis of the Speech Records of the US Congress and the Japanese Diet

Hiroki Takikawa[1] and Takuto Sakamoto[2,*]

[1]Frontier Research Institute for Interdisciplinary Sciences (FRIS), Tohoku University, Sendai, JAPAN
[2]Graduate School of Arts and Sciences, University of Tokyo, Tokyo, JAPAN
[*]Corresponding author: sakamoto@hsp.c.u-tokyo.ac.jp

April 23, 2017

*Keywords: moral psychology, moral foundation theory, sentiment analysis, natural language processing, comparative analysis*

There has been growing interest in moral-emotional phenomena in psychology, sociology and other social sciences. One of the main issues in this field of study centers on the relationship between public/political discussion and moral-emotional framework. The first major attempt was done by J. Haidt(2012), a leading advocate of 'moral psychology'. He argues that there are five distinct moral foundations in our moral framework ("care/harm", "fairness/reciprocity", "ingroup/loyalty", "authority/respect" and "purity/sanctity") and that the emphasis on each moral foundation differs depending on people's political ideology. According to Haidt, liberals endorse the first two foundations while conservatives evenly draw on all of the five foundations. His findings were originally based on psychological experiments and questionnaire surveys, but in recent years researchers including Haidt and his colleagues themselves (Graham et al. 2009) have been studying "moral behavior 'in the wild'"(Dehghani et al. 2016) by utilizing the computational linguistic analysis of large-scale corpuses created from social media data and other sources.  A similar development can be seen in the closely related field of sentiment analysis of political discussion. Here, the use of "big data" (Wojcik et al. 2015) has led to another intriguing finding: conservatives display happiness (positive emotion) *less* than liberals, which directly opposes the conventional wisdom in the field. However, these arguments, which have mostly been advanced in the social context of the United States, lack a comparative perspective. This might cause serious problems because, as sociologists argue, moral-emotional phenomena are deeply embedded in social and institutional environments. An argument based on just a single country's case involves a great risk of making erroneous generalizations.

As the first step towards overcoming these limitations, we compare the emotional and moral structures of political and public discussion observed in the U.S. and Japan by employing extensive text data that cover these two countries. More specifically, we conduct sentiment analysis and moral foundation analysis of floor debate in the U.S. Congress and the Japanese Diet over a long period of time. In so doing, we reexamine the moral foundation theory proposed by Haidt as well as the somewhat controversial relationship between emotions and political ideology suggested by Wojcik et al. In drawing comparisons between U.S. and Japan, we pay particular attention to institutional differences between the two legislative bodies (e.g., the extent of party restrictions on votes; the degree of party confrontation, etc.). We investigate whether and how these institutional differences might affect the moral-emotional structure of political discussion in different social settings.

Data and Methods

For each of the two countries, we obtained a large volume of text data that record floor deliberations among legislators over a long period of time. Following several steps of natural language processing procedure (tokenizing sentences, removing stopwords, applying morphological analysis…), we converted the speech data into a 1-gram bag of words. Against these corpora obtained from the U.S. and the Japanese legislative bodies, we conducted dictionary-based sentiment and moral analyses. That is, we counted frequencies of relevant terms in the corpora of each country with the help of well-established sentiment and moral dictionaries.

For the U.S. speech data, we relied on an online version of *Congressional Records*. We used data from both the House of Representatives and the Senate during a period of 1994-2016. The data size amounts to 6,423,453 remarks. In the case of Japan, we obtained the minutes of the plenary sessions and the budget committees of the House of Representatives and the House of Councilors from 1947 to 2016, using the Diet Conference Proceedings, which is also available online. The number of remarks is 1,608,731 in total. As to the dictionaries, we employed the moral-term dictionary developed by Haidt and his colleagues, and its Japanese translation (by ourselves), for moral foundation analysis of the U.S. and the Japanese datasets. Regarding sentiment analysis, we chose to use separate, equally well-established, dictionaries for the two countries: positive- and negative-emotional English words found in LIWC 2015 (Pennebaker et al. 2003) and Japanese Sentiment Dictionary (Volume of Nouns) (Higashiyama et al. 2008).

Results and Discussion

The analyses described above revealed more varied and more nuanced relationships between public/political discussion and moral-emotional framework than the preceding arguments had suggested. These relationships, moreover, strongly indicates the possible causal effects of institutional differences between American and Japanese parliamentary politics.

(1) Sentiment analysis

Analysis of the US congressional debate suggests the following relationships between political ideology and expressed emotions. First, Democrats (assumed to be liberals) tend to express the positive emotions more frequently than Republicans (assumed to be conservatives) do, which is in line with the argument of Wojcik et al (2015). Second, however, negative emotions also tend to be more strongly expressed by Democrats, posing a significant caveat to their findings (see Fig.1). Moreover, if we look at Japan, a significantly different pattern can be seen. In the Japanese diet debate, there is a clear tendency for members of LDP (Liberal Democratic Party), a conservative, long-time ruling party, to express positive emotions more, and negative emotions less, than liberal SDP (Social Democratic Party) members do (Fig.2). We suspect that this tendency comes from the highly stratified and confrontational parliamentary politics that has characterized the post-war Japanese democracy.

(2) Moral foundation analysis

In Haidt's moral dictionary, the words assigned to each of the five moral foundations are further classified into virtue (positive) and vice (negative): the former means "foundation-supporting words" such as "kindness, patriot and obey", while the latter corresponds to "foundation-violating words such as "hurt, betray and disrespect". This distinction is the key to understanding significant differences found between moral frameworks employed in the U.S. and Japan.

The analysis of the U.S. data shows that the Democrats are more concerned about "harm and care" in both positive and negative senses than Republicans do, which is consistent with the moral foundation hypothesis. Otherwise, however, our extensive analysis did not give support to this hypothesis. For example, Republicans, rather than Democrats, express more interests in "fairness/reciprocity" and less interests in "ingroup/loyalty" (Fig.3). Furthermore, the analysis of the Japanese data reveals unexpected but intriguing patterns in moral framework. Among others, liberals (SPD) in Japan have consistently shown more *negative* moral concerns in all five foundations than conservatives (LDP) (Fig.4). This indicates that in more polarized Japanese political settings, dominant conservative arguments have been 'counteracted' by strongly negative moral concerns of liberals. It again suggests the possible causal influences of distinct social and institutional environments on the characteristics of moral-emotional structure of political discussion, although the exact mechanism should be clarified in the future work.

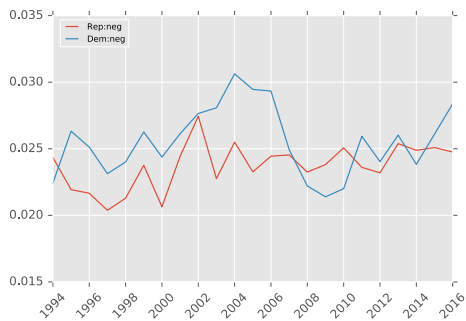

Fig.1: frequencies of negative-emotion words (US)

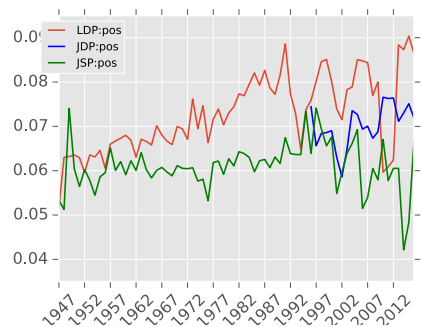

Fig.2: frequencies of positive-emotion words (JP)

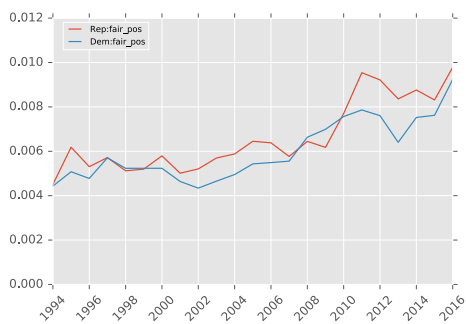

Fig.3: frequencies of positive-fairness words (US)

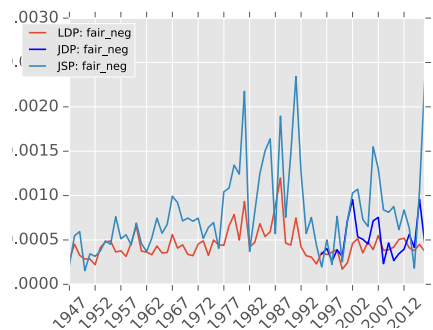

Fig.4: frequencies of negative-fairness words (JP)